\documentclass{aa}
\usepackage{graphics}
\begin{document}

   \thesaurus{06     
	      (02.01.2;	 %
		   08.02.1;
		   08.09.2;
		   13.25.5)}

   \title{Disk mass accretion rate and infrared flares in GRS~1915+105}

   \author{Belloni, T.\inst{1}, Migliari, S.\inst{1,2}, Fender, R.P.\inst{3}
	  }

   \offprints{T. Belloni}

   \institute{Osservatorio Astronomico di Brera, Via. E. Bianchi 46, I-23870,
	Merate, Italy
	e-mail: belloni@merate.mi.astro.it
	 \and
	      Dipartimento di Fisica, Universit\`a degli Studi di Milano,
			Via Celoria 16,
			I-20133, Milano, Italy
      e-mail: migliari@merate.mi.astro.it
	 \and
	      Astronomical Institute ``Anton Pannekoek'' and Center
		for High-Energy Astrophysics, 
		University of Amsterdam,
		Kruislaan 403,
			1098 SJ, Amsterdam, The Netherlands
	e-mail: rpf@astro.uva.nl
	     }

   \date{Received ; accepted 15 May 2000}

   \maketitle

   \begin{abstract}

We have analyzed in detail a set of Rossi X-ray Timing Explorer (RXTE)
observations of the galactic microquasar GRS~1915+105 corresponding to
times when quasi-periodic oscillations in the infrared have been
reported. From time-resolved spectral analysis, we have estimated the
mass accretion rate through the (variable) inner edge of the accretion
disk. We compare this accretion rate to an estimate of the mass/energy
outflow rate in the jet.  We discuss the possible implications of
these results in terms of disk-instability and jet ejection, and in
particular note an apparent anti-correlation between the accretion and
ejection rates, implying that the gas expelled in the jet must leave
the accretion disk before reaching its innermost radius.

   \end{abstract}

%

\section{Introduction}

GRS~1915+105 is a transient X-ray source discovered in 1992 with WATCH
(Castro-Tirado, Brandt \& Lund 1992). Since then it has probably never
switched off completely and it has remained as a highly variable
bright X-ray source (see Sazonov et al. 1994; Paciesas et al. 1996;
Bradt et al. 2000).  It is the first Galactic object that was found to
show superluminal expansion in the radio (Mirabel \& Rodr\'\i guez
1994). The interpretation of this phenomenon in terms of relativistic
jets (Rees 1966) implies bulk velocities of the ejecta of $\geq 0.9c$
at an angle of 60--70 degrees to the line of sight (Mirabel \& Rodr\'\i
guez 1994, Fender et al. 1999, Rodr\'\i guez \& Mirabel 1999).
Because of the high value of the extinction on the line of sight, no
optical counterpart is available, but an infrared counterpart has been
found (Mirabel et al. 1994).  The source is suspected to host a black
hole because of its high X-ray luminosity and its similarity with
another Galactic superluminal source GRO~J1655-40 (Zhang et
al. 1994), for which a dynamical estimate of the mass is available
(Orosz \& Bailyn 1997).

Four years of monitoring with the All-Sky Monitor (ASM) on board RXTE
showed that the 2-10 keV flux of GRS~1915+105 is extremely variable,
considerably more than any other known X-ray source (see Bradt et al. 2000).
See Belloni et al. (2000) for a complete reference list of RXTE
observations of the source.

Belloni et al. (1997a,b), from the analysis of selected X-ray spectra,
showed that the X-ray variability of the source can be interpreted as
the repeated appearance/disap\-pea\-rance of the inner portion of the
accretion disk, caused by a thermal-viscous instability. During the
low-flux intervals, when the source spectrum hardens considerably, the
inner disk up to a certain radius becomes unobservable and is slowly
re-filled again.  A more complete picture of these variations, where
the observations were classified into twelve different classes and
another type of (soft) low-flux intervals was presented, was shown by
Belloni et al. (2000).  Additional spectral analysis has been
presented by Markwardt et al. (1999) and Muno et al. (1999), who
analyzed in detail the connection between QPOs and energy spectra in
GRS~1915+105.  One of the problems caused by the exceptional
variability of the source is that it is difficult to estimate the
accretion rate through the disk or even to rate observations according
to accretion rate.

Quasi-periodic variability in the radio, infrared and millimetre bands
has been discovered (Pooley 1995, Pooley \& Fender 1997; Fender et
al. 1997; Fender \& Pooley 2000).  Fender et al. (1997) suggested that
these oscillations could correspond to small ejections of material
from the system.  Indeed, these oscillations have been found to
correlate with the disk-instability as observed in the X-ray band
(Pooley \& Fender 1997; Eikenberry et al. 1998,2000; Mirabel et al. 1998).
This suggests that (some of) the gas is
ejected from the inner disk during each low-flux interval.
On longer time scales an analogous pattern is
observed in the form of major relativistic ejections occurring at the
end of a 20-day X-ray dip or `plateau' (Fender et al. 1999).

\begin{figure}
\resizebox{\hsize}{!}{\includegraphics{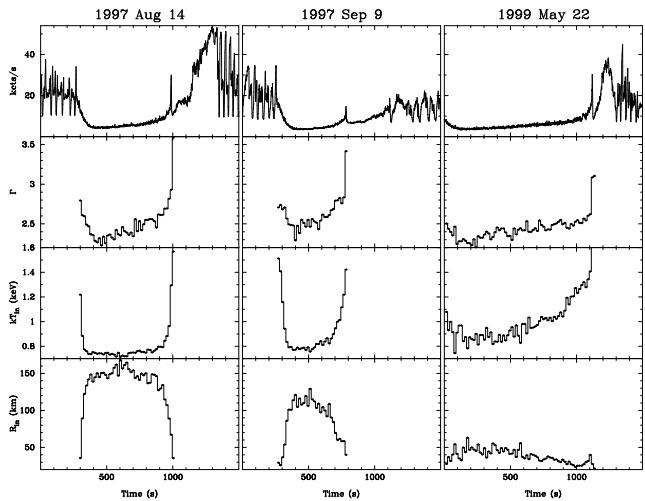}}
\caption{PCA light curves and corresponding timing evolution of
        selected spectral parameters 
        (power-law photon index $\Gamma$, inner disk temperature kT$_{\rm in}$ 
         and
         radius R$_{\rm in}$) from the three
	observations for which detailed analysis was possible (see text).
	Light curves have a 1s time resolution, parameters are from 16s
	bins. The parameters are shown only for the state C intervals
	(see Belloni et al. 2000).}
\label{fig1}
\end{figure}

In this Letter we present the results of detailed time-resolved
spectral analysis of RXTE/PCA data of observations when
(quasi-)simultaneous infrared data are available. We estimate the
value of the accretion rate through the disk for each observation and
show that it is anticorrelated with the estimated jet power.

\section{Data analysis}

The published infrared observations of GRS~1915+105 for which there
are simultaneous or quasi-simultaneous (ie. within 2 days) RXTE/PCA
data are those from Mirabel et al. (1998), Eikenberry et al. (1998),
Fender et al. (1998), Eikenberry et al. (2000), Fender \& Pooley,\
(2000). All observations reveal very variable X-ray light curves (see
Table 1), corresponding to classes $\beta$, $\nu$ and $\theta$ in the
classification by Belloni et al. (2000).

\begin{table*}
      \caption[]{Log of PCA observations and summary of spectral parameters
		 (see text). Classes in column 3 correspond to the
		classification from Belloni et al. (2000).}
\begin{flushleft}
\begin{tabular}{lcccccccc}
     Date		  &
     Obs\#		  &
     Class		  &
     T$_{\rm start}$(UT)	  &
     $\Delta$ t (s)	  &
     R$_{\rm max}$(km)	  &
     $\dot{M}_{\rm disk}$ (M$_\odot$/yr)	   &
     $\dot{M}_{\rm J}	    $ (M$_\odot$/yr)	   &
     $P_{\rm J}$ (erg s$^{-1}$)\\
\hline
14/8/97 & 20186-03-03-01 &$\beta$&  4:02   & 530-690	  & 170$\pm$
14	  &1.3$\times 10^{-7}$& 6$\times 10^{-7}$(a) & $9\times 10^{37}$\\
09/9/97 & 20402-01-45-03 &$\beta$&  6:00   & 500-720	  & 128$\pm$
13	  &7.1$\times 10^{-8}$& 3$\times 10^{-7}$(b) & $1\times 10^{38}$ \\
15/9/97 & 20186-03-02-00 &$\theta$& 12:31  & 600-1000	  &  ---$^c$
&---$^c$ &5$\times 10^{-7}$(d*) & $9\times 10^{37}$ \\
10/7/98 & 30182-01-03-00 &$\nu$&  5:05	   & 2250-3500$^e$& 288$\pm$
27	  &2.7$\times 10^{-7}$&$10^{-7}$(f) & $4 \times 10^{37}$\\
22/5/99 & 40702-01-02-00 &$\nu$& 20:41	   & 1100-1370	  & 55$\pm$ 13
&8.0$\times 10^{-9}$&2$\times 10^{-6}$(g*) &  $3 \times 10^{38}$\\
\end{tabular}
\end{flushleft}
    \begin{list}{}{}
	 \item[$^{\mathrm{a}}$] from Eikenberry et al. (1998);
	       $^{\mathrm{b}}$ from Mirabel et al. (1998);
	       $^{\mathrm{c}}$ not measurable;
	       $^{\mathrm{d}}$ from Fender \& Pooley (1998)
	 \item[$^{\mathrm{e}}$] determined from IR data;
	       $^{\mathrm{f}}$ from Eikenberry et al. (2000);
	       $^{\mathrm{g}}$ from Fender \& Pooley (2000);
	       $^{\mathrm{*}}$ quasi-simultaneous
     \end{list}
\end{table*}

For each observation, we produce light curves at 1s time resolution
(from {\tt Standard1} data) and isolated the long hard low-flux
intervals corresponding to state C (unobservable inner disk) of
Belloni et al. (2000). For each interval, we measured its length from
the light curve (see Table 1).  Then we accumulated spectra on a time
scale of 16 seconds from {\tt Standard2} data, thus retaining the full
energy resolution and coverage of the PCA. From each spectrum, we
subtracted the background estimated with {\tt pcabackest}
vers. 2.1b. We did not correct for deadtime effects, but we do not
expect this effect to be too important. For each observation in PCA
epoch 3 we produced a detector response matrix using {\tt pcarsp},
while for epoch 4 we used the response provided on line by K. Yahoda
\footnote{\tt http://lheawww.gsfc.nasa.gov/users/keith/epoch4/}.  
We fitted
each spectrum with the ``standard'' model used for black-hole
candidates, consisting of the superposition of a multicolor
disk-blackbody and a power law. By assuming a distance of 12.5 kpc and
a disk inclination of 70$^{\circ}$ (Mirabel \& Rodr\'\i guez 1994), we
can derive from the fits the inner radius of the accretion disk.
Correction for interstellar absorption (fixed to $6\times
10^{22}$cm$^{-2}$, see Belloni et al. 2000) and an additional emission
line (fixed at 6.4 keV) were also included. A systematic error of 1\%
was added. The value of the reduced $\chi^2$ was usually around 1,
although some fits were slightly worse. The resulting interesting
parameters (inner disk radius and temperature, slope of the power law)
as a function of time are shown in Fig. 1 for three of the five
observations, for which this automated procedure gave good results.
The remaining two observations had to be treated more carefully. The
observation from 1997 Sep 15th, the only one from class $\theta$,
resulted in an extremely strong power law component, with a photon
index steeper than 3.  The softness and intensity of this component
made it impossible to obtain sensible values for the disk parameters,
although there is evidence of its presence. This enhanced power law is
probably the reason of the difference between this class and the
others (see Belloni et al. 2000). The observation from 1998 July 10th
did not include full state-C intervals: in this case, we measured the
length of the intervals from the infrared (Eikenberry et
al. 2000). Also, the inner disk radius resulted to be larger and
therefore more difficult to measure as this component is softer. In
order to estimate the disk parameters, we produced a 32s spectrum
corresponding to the bottom of the dip only and obtained the best fit
parameters, corresponding to the largest inner radius.  This is the
reason why there is only one point for this observation in Fig. 2.

\begin{figure}
\resizebox{\hsize}{!}{\includegraphics{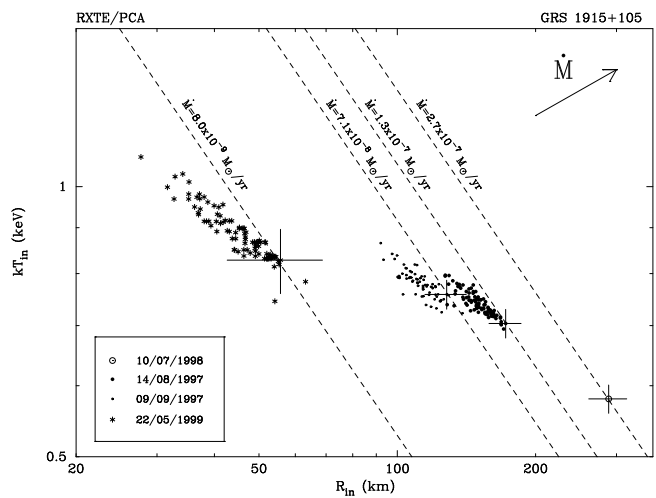}}
\caption{Evolution of temperature at the inner disk radius versus inner
	disk radius for the four observations for which a reliable
	estimate could be obtained (see text). Only the low-flux sections
	of the data from Fig.1 are shown. Typical errors are shown for each
	observation. The dashed lines correspond to different values of
	disk accretion rate according to the thin disk model.}
\label{fig2}
\end{figure}

\section{Results}

In principle, from each spectrum the accretion rate through the
measured inner radius of the disk could be measured from the values of
kT$_{\rm in}$ and R$_{\rm in}$ (see Belloni et al. 1997a) by using the
expression from a standard thin accretion disk. However, given the
errors on these parameters, this measurement is too uncertain.  In
order to obtain an improved estimate of the disk accretion rate or,
better, a ranking of the observations in terms of accretion rate
(since the actual values of the inner disk radii obtained with the
multicolor disk-blackbody model are probably underestimates, see
Merloni, Fabian \& Ross 1999), we plotted the values corresponding to
the deepest parts of the X-ray light curves in a kT$_{\rm in}$
vs. R$_{\rm in}$ plane (see Fig. 2). If for each observation the disk
accretion rate was constant, the points should lie on the diagonal
lines corresponding to a slope $-$3/4 (as, for a given $\dot{M}$, $T
\propto R^{-3/4}$ -- Belloni et al. 1997a).  Their actual distribution
is flatter, showing that there is a deviation from the expected law,
but it is interesting to note that the distributions lie on parallel
curves in the log-log plane. This indicates different values of the
disk accretion rate. Lines corresponding to the larger measured radius
for each of the four observations are shown in Fig. 2 with their
associated accretion rate value. Typical 1$\sigma$ errors are also
shown.  Although the actual values for the accretion rate are probably
not accurate, on the basis of this plot we can rank the observations
by accretion rate.  It is important to note that the accretion rate
measured this way correspond to matter passing {\it through} the
observed inner radius of the disk only: if some matter leaves the disk
before that radius, its presence cannot be detected with this
procedure.  This estimate of accretion rate can be double checked by
considering the length of the state C intervals, which Belloni et
al. (1997a,b) interpreted as the viscous time scale of the disk at the
edge of the unobservable region which is refilled. The observation
from 1999 May 22nd has a smaller inner disk radius (see Fig. 2) than
the 1997 ones and a longer re-fill time (Tab. 1), indicating a lower
value of the accretion rate. The 1998 July 10th observation has a much
larger inner disk radius than the 1997 ones, by a factor of 1.7 and
2.3, which would correspond to a re-fill time longer by a factor 6.4
and 18 respectively, while it is much shorter, indicating a higher
accretion rate.

\section{Discussion}

The results of our analysis indicate that, at least for observations
of class $\nu$ and $\beta$ (which have many similar traits), we have a
way to estimate the disk accretion rate during an instability event,
when the inner disk radius grows from its ``minimum'' value of
$\sim$30 km and slowly moves back to it. Although we know that the
measured value is only an underestimate, it is natural to associate
this minimum value with the innermost stable orbit.  It is interesting
to compare these values, or at least their ranking, with the rate of
ejection in the jets.  As we mentioned above, the accretion rate
measured through this procedure is associated to matter flowing {\it
through} the observable inner edge of a geometrically thin accretion
disk.  
Some of the accreting gas must leave the accretion disk to form the jet,
unless it is entirely composed of pairs generated by photon-photon
interactions.
and how this happens is basically unknown. 
There are two extreme possibilities: either matter
ejected in the jet leaves the accretion disk before entering the innermost
regions, thus not contributing to our measured disk accretion rate
(case 1), or it leaves it after passing through our measured inner
disk radius, in which case it is a fraction of the accretion rate we
measure (case 2).  In case 1, if the fraction of matter in the jet is
constant and the total external accretion rate (disk+jet) is variable,
we expect a positive correlation between disk accretion rate (from X
rays) and disk ejection rate (from the infrared). If the fraction is
variable and the total is constant, these quantities should be
anticorrelated.  In case 2, if the fraction of matter in the jet is
constant, we expect a positive correlation, while the constant total
is in this case not possible as the total would be what we measure, which is
not observed to be constant.  If both fraction and total
vary, the situation is complicated. Of course, there is a spectrum of 
intermediate possibilities, where the jet production is connected to the
inner region of the disk in a way that would not allow to dissociate the
two processes. With the paucity of existing data, we limit ourselves to the
extreme cases. Notice that measuring an
anti-correlation would be an indication against case 2. 

Table 1 also lists an estimate of the mass ejection rate
$\dot{M}_{\rm J}$. This is based upon an equipartition calculation for one
proton for each electron, negligible kinetic energy associated with
the repeated ejection events, and an average over the repetition
period of the oscillations.  Note that there is a systematic
uncertainty in these numbers due to lack of knowledge of the intrinsic
electron spectrum which corresponds to the observed flat-spectrum
radio--infrared emission. However, unless the spectral form of the
distribution changes between observations then the effect is the same
for all data sets and the ranking remains the same. Of course we may
be observing synchrotron emission from a pair plasma with no baryonic
content, in which case the amount of power being supplied to the jet,
$P_{\rm J}$, makes more useful comparison with the accretion rate; this
value is also listed in Table 1.  For more details of how these
quantities are calculated, see Fender \& Pooley (2000).  Either way,
there appears to be an {\em anticorrelation} between accretion rate
inferred from the X-ray spectral fits and the outflow rate of
mass/energy in the jet.  The low number of points in our sample
prevents us from saying something more firm.  Notice that an
anticorrelation is also suggested by the strong flat-spectrum radio
emission observed during long `plateau' intervals; periods when
Belloni et al. (2000) estimate that the accretion rate must be very
low.  We also note that the faint infrared flares reported by
Eikenberry et al.  (2000) do not appear to be different from the
others in other respects, as the X-ray light curves are too
undersampled to allow a detailed correlation.

If future observations show that disk accretion rate and jet ejection
rate are indeed anti-correlated, the following scenario could be
speculated. A fraction of the accreting gas leaves the geometrically
thin accretion disk before reaching the inner edge (from which it
would fall into the black hole) and goes into a hot corona. The details
are not known, but our results indicate that this does not happen
after the inner edge. As the
disk refills, the inner radius moves inwards, more soft photons from
the disk reach the corona, which causes its Comptonization emission to
soften gradually. At the end of the instability period, when the disk
is refilled down to the innermost stable orbit, this ``reservoir'' of
hot gas is expelled to produce the jet, resulting in the observed
infrared / mm / radio emission, causing the power-law component to
steepen dramatically and to cause the sudden change in the X-ray count
rate and spectral parameters.  Notice that, as we remarked earlier,
the distributions of points in Fig. 2 are flatter than the expected
curve for a constant disk accretion rate according to a standard thin
disk: in other words, as the inner disk radius decreases, the disk
accretion rate seems to decrease as well. This could mean that the
process that re-routes some gas from the disk to the corona becomes
more efficient closer to the central object, and therefore the
fraction of matter going into the corona increases as the disk
refills.


%
\begin{acknowledgements}
      We thank G. Ghisellini and M. Tagger 
for useful discussions.
\end{acknowledgements}

\end{document}